\documentclass[pdftex,twocolumn,epjc3]{svjour3}          

\RequirePackage[T1]{fontenc}

\smartqed  

\RequirePackage{graphicx}
\RequirePackage{mathptmx}      
\RequirePackage{flushend}
\RequirePackage[numbers,sort&compress]{natbib}
\RequirePackage[colorlinks,citecolor=blue,urlcolor=blue,linkcolor=blue]{hyperref}

\journalname{Eur. Phys. J. C}

\usepackage{amsmath}
\def\abs#1{\lvert #1\rvert}
\begin{document}

\title{Physical model of dimensional regularization}

\author{Jonathan F. Schonfeld\thanksref{e1,addr1}}

\thankstext{e1}{e-mail: jschonfeld@aya.yale.edu}

\institute{Lexington, Massachusetts USA 02420\label{addr1}}

\date{Received: date / Accepted: date}

\maketitle

\begin{abstract}
We explicitly construct fractals of dimension $4{-}\varepsilon$ on which
dimensional regularization approximates scalar-field-only quantum-field-theory
amplitudes.  The construction does not require fractals to be Lorentz-invariant
in any sense, and we argue that there probably is no Lorentz-invariant fractal
of dimension greater than $2$.  We derive dimensional regularization's
power-law screening first for fractals obtained by removing voids from
$3$-dimensional Euclidean space.  The derivation applies techniques from
elementary dielectric theory.  Surprisingly, fractal geometry by itself does
not guarantee the appropriate power-law behavior; boundary conditions at
fractal voids also play an important role.  We then extend the derivation to
$4$-dimensional Minkowski space.  We comment on generalization to non-scalar
fields, and speculate about implications for quantum gravity.
\end{abstract}

\section{Introduction}\label{sec1}

Is ``dimension deficit'' really the correct physical meaning of the parameter
$\varepsilon$ in dimensional regularization?  The only way to prove it is by
explicitly constructing a fractal spacetime on which dimensional regularization
approximates quantum-field amplitudes.  Introducing such a construction for
scalar-only quantum field theories is the purpose of this paper.

Ideally, this is the first step in a longer research program aimed at extending
this construction to non-scalar fields.  Even if that doesn't materialize,
the scalar construction should be of interest in its own right, as it casts
a fresh light on the foundations of dimensional regularization, one of the
cornerstones of modern quantum field theory.  After all, other schemes such
as Pauli--Villars and lattice regularization have well-defined physical
meanings that enable scientists to benefit from intuition established in a
variety of other domains.  Why not dimensional regularization?

The remainder of this paper is organized as follows.  Section~\ref{sec2}
presents essential concepts and argumentation about dimensional regularization
and fractals, and sets the stage for the constructions and derivations that
follow.  Section~\ref{sec3} derives the power-law screening characteristic of
dimensional regularization for propagators on fractals defined by removing
voids from $3$-dimensional Euclidean space, in order to establish basic
intuition and lines of argumentation.  The results of Section~\ref{sec3}
apply techniques from elementary dielectric theory.  Surprisingly, fractal
geometry by itself does not guarantee the power-law behavior required
for dimensional regularization; boundary conditions at fractal voids also
play an important role.  Section~\ref{sec4} extends Section~\ref{sec3} to
fractals in $4$-dimensional Minkowski space.  Note that in Section~\ref{sec4}
the fractals themselves are not Lorentz invariant, but that's alright because
(see below) anisotropy in fractal power-law scaling appears to have no impact
on dimensional regularization for small $\varepsilon$.  (We will in fact
argue that there is no such thing as a relativistically invariant fractal
with dimension greater than $2$.)  Section~\ref{sec5} contains a discussion
of weaknesses in our reasoning, as well as prospects for generalization to
non-scalar fields, and speculation about implications for quantum gravity.

\section{Preliminaries about dimensional regularization and
fractals}\label{sec2}

Dimensional regularization \cite{hoof72} for scalar fields amounts to
changing the momentum-space volume element $d^4p$ in Feynman diagrams to
$\abs{p/\mu}^{-\varepsilon}d^4p$, where $\abs{p}$ is the Minkowski norm of momentum $p$,
$\varepsilon$ is positive and $\mu$ is a fixed scale.  The important thing is
that the multiplier behaves like a fractional power of the scale factor
as $p$ scales to infinity along any fixed direction.  That is what turns
logarithmic divergences in $p$ into poles in $\varepsilon$.  This has the same
effect as multiplying the scalar propagator (instead of the integration
volume) in momentum space by $\abs{p/\mu}^{-\delta}$, where $\delta =\varepsilon/2$, because
the only divergent loops have two scalar propagators.  For this reason,
the goal of the fractal constructions that follow is to show that scalar
propagators in fractal spacetimes exhibit screening of the form $\abs{p/\mu}^{-\delta}$
in momentum space for large momentum $p$ or, as appropriate, $\abs{x\mu}^{+\delta}$ in
position space for small position $x$ and non-negative $\delta$.  (More precisely,
when quantum amplitudes are defined by path integrals over random fractals
[see below] obtained by removing voids from linear spacetime, then the
quantum amplitudes, when ensemble-averaged over random fractals, numerically
correspond to Feynman diagrams in the underlying integer-dimensional linear
spacetime with propagators screened as described above.){\emergencystretch7pt\par}

There is a hitch, however.  In section~\ref{sec4} we shall find ourselves dealing with
fractals that are not themselves Lorentz-invariant.  This means that we will
really show that propagators in position space scale as $\abs{x\mu f(\Omega)}^{+\delta}$ at
short distance or $\abs{g(\Omega)p/\mu}^{-\delta}$ at large momentum, where $\Omega$ is solid
angle in four dimensions and $f$ or $g$ is some function that's nonzero almost
everywhere.  But that's alright as far as Lorentz-invariance of quantum
amplitudes is concerned, because the function $f$ has no \emph{material} impact on
dimensional regularization for small dimension deficit:  Small $\varepsilon$ (or
$\delta$) ensures that integration over $\Omega$ doesn't diverge; and ignoring terms of
order $\varepsilon$ and higher ensures that the only quantitative effect that
$g$ has on Feynman integrals is to modify the effective value of $\mu$, because
the ${\rm O}(\varepsilon)$ term in $g(\Omega)^{-\varepsilon}$ can only manifest itself by
multiplying the $1/\varepsilon$ linear-scale divergence by the integral of
$\ln[g(\Omega)]$ over all solid angles.

The constructions in this paper focus on random ``take-away'' fractals.
Randomness exempts us from the complications of accidental crystallographic
symmetries.  For the purposes of this paper, a random ``take-away''
fractal is a set formed by the following recursive procedure: Start with
a linear space of integer dimension $D$, and a reference void of volume $V$.
Distribute points randomly throughout space with arbitrary density $\rho$, and,
centered at each such point, remove a copy of the reference void.  Call this
the zeroth iteration.  Now choose an arbitrary scale factor $\xi>1$ and define
the $k$'th iteration inductively as follows:
\begin{itemize}
\item Distribute points randomly with density $\rho\xi^{Dk}$ throughout
whatever part of the Euclidean space has not been removed by preceding
iterations.
\item Centered at each such point, remove from the $k-1$'st iteration a copy
of the reference void linearly scaled by factor $\xi^{-k}$.
\end{itemize}
In the limit of infinite $k$, what's left has fractal dimension
$D+\ln(1-\rho V)/\ln\xi$ \cite{mand}.  The factor $(1-\rho V)$ is the volumetric proportion
of iteration $k-1$ removed by iteration $k$, and the ratio of logarithms is
clearly minus a dimension deficit for physical $\rho V<1$. (The geometry of the
reference void can have its own probability distribution, but this is more
generality than we require.  Also, for small $\rho V$ it is unimportant that
voids removed at iteration $k$ might overlap with one another or with voids
removed at earlier iterations.)

A \emph{relativistic} fractal would have a Lorentz-invariant reference void.
But such a void -- bounded by hyperbolic spheres -- would have infinite $V$.
Interestingly, if the starting linear space had dimension $D=2$, a fractal
could still be defined.  For example, consider as a reference void the space
between a forward light cone and a mass shell.	Basically this is a central
lobe flaring into two wings whose thickness falls like one over the distance
from the vertex.   A random distribution of such voids produces a fractal
after just a single iteration, because there are $\rho2\pi rdr$ void centers at
distance $r$ from any reference point, and their wing width falls like $1/r$
(basically, $1/r$ is analogous to $V\xi^{-Dk}$ in the fundamental fractal
definition, and $\rho2\pi r$ is analogous to $\rho\xi^{Dk}$). For arbitrary $D$ the
number of void centers at distance $r$ is proportional to $\rho r^{D-1}dr$ but wing
(really a cup lip) width still falls like $1/r$, so the two factors match only
for $D=2$ and we conclude that there probably are no relativistic fractals in
higher dimensions.  This reinforces the importance of our earlier discussion
about Lorentz invariance vs.\ nontrivial $f$ or $g$.  Perhaps this uniqueness
of $D=2$ is behind the tendency of renormalization-group flows to converge to
$2$-dimensional fractals in reduced models of gravity \cite{laus}. [Alternatively, 
one might construct a relativistic fractal by replacing the idea of a single 
infinite-volume Lorentz-invariant reference void with an appropriately weighted 
ensemble of voids created by applying Lorentz boosts to a single finite-volume 
``seed'' void.  But one is driven to the same conclusion about $D = 2$ because the 
Lorentz-invariant measure on the set of all boosts has itself infinite total weight.]

\section{Propagator on fractal derived from Euclidean 3-space}\label{sec3}

As indicated above, the narrow mathematical objective of this paper is
to derive power-law screening at small distances or large momenta for
wave-equation propagators in $4$-dimensional Minkowski space limited by a
fractal distribution of voids.	To make the thought process as clear as
possible, we build to this objective with three cases of successively
increasing sophistication.  The last case is Minkowski space.

For the first case, consider recovering the $\ln r$ Green's function for potential
theory in two dimensions by limiting three dimensions to the space between
two closely-separated parallel planes.	In school we encounter the problem
of a point charge between two parallel conducting planes, but because the
infinite sequence of image charges involves alternating signs, the potential
does not approach $\ln r$ for vanishing plane separation \cite{pump}.  If instead of
conducting planes -- i.e.\ constant-value Dirichlet boundary condition
-- we impose the other canonical potential-theory boundary condition
-- zero-normal-derivative Neumann -- the image charges are in the same
locations but all have identical sign.	So they add coherently to produce
$\ln r$ for vanishingly small plane separation.  Naively, the coefficient of
$\ln r$ diverges as $q/a$, where $q$ is the original point charge and $a$ is plane
separation, but $a$ cancels out because the 2D Green's function is meant to be
integrated over the limiting plane, while in 3D it's to be integrated over
the space between the converging planes, and that volume is proportional to $a$.
This sets a pattern for the cases that follow: invocation of Neumann boundary
conditions modulated by vanishing volume between voids.

For the second case, consider the Green's function at short distance for
potential theory in $D=3$-dimensional Euclidean space limited by a fractal
distribution of spherical voids. If the spheres are small, the field around
each primarily induces an electrostatic dipole \cite{jack} with polarizability $\gamma_k$
for the spheres of iteration $k$.  According to dielectric theory \cite{jack}, these
spheres collectively amplify or shield a distant charge by a factor
\begin{equation}
\Phi_k=\left[1+\frac{4\pi\rho\xi^{3k}\gamma_k}{1-\frac{4\pi}{3}\rho\xi^{3k}\gamma_k}\right]^{-1}.
\label{eq1}
\end{equation}
Each iteration of the fractal process multiplies the Green's function
(potential) of a point charge by this factor in the space between spheres,
\emph{but only for iterations whose spheres are smaller than the distance to
the point charge}, since larger spheres don't fit.  At the same time,
each iteration also multiplies the point-charge potential by a factor of
$(1-\rho V)$ for integration volume regardless of sphere size.  In other words,
the Green's function for point charge $q$ becomes
\begin{equation}
-\frac{q}r \prod_{k=0}^{\infty}(1-\rho V) \Phi_k  \prod_{l=0}^{l_{\max}}\Phi_l^{-1}
\label{eq2}
\end{equation}
where $l_{\max}$ is the highest iteration whose spheres are larger than or equal
to $r$.  For the infinite product to be well-defined, $(1-\rho V)\Phi_k$ must be unity.
Thus we discover that spheres have to come in a mix of boundary conditions
so that on average
\begin{equation}
\frac{4\pi}3 \rho\xi^{3k} \gamma_k=\frac{-\rho V}{3-\rho V}.
\label{eq3}
\end{equation}

It is elementary to show that polarizability for a spherical void at iteration
$k$ is $3V/4\pi\xi^{3k}$ for Dirichlet boundary conditions and $-3V/8\pi\xi^{3k}$
for Neumann.  So Eq.~\eqref{eq3} says that for every iteration the voids must be
a mix of $2(4-\nobreak\rho V)/3(3-\rho V)$ Neumann and $(1-\rho V)/3(3-\rho V)$ Dirichlet.
As a result, expression \eqref{eq2} reduces to
\begin{equation}
-\frac{q}r (1-\rho V)^{l_{\max}}.
\label{eq4}
\end{equation}
Since $l_{\max}$ satisfies $r\sim$ radius of iteration-$l_{\max}$ sphere, proportional to
$V^{1/3}/\xi^{l_{\max}}$, expression \eqref{eq4} amounts to power-law screening of the form
\begin{equation}
\left(\frac{r}{V^{1/3}}\right)^{\!-\ln(1-\rho V)/\ln\xi}.
\label{eq5}
\end{equation}
The exponent in expression \eqref{eq5} is the dimension deficit.

\section{Propagator on fractal derived from Minkowski 4-space}\label{sec4}

In Minkowski space, we must step away from fractals defined by spherical voids
because the wave equation -- rather than Poisson's equation -- prevails.
The $4$-space wave equation is governed by initial conditions on $3$-dimensional
space and boundary conditions on $2$-dimensional walls, in contrast with the
$4$-space Poisson equation, which would require conditions on the entirety of
arbitrarily shaped $3$-di\-men\-sion\-al boundaries.  For this reason we now assume
cylindrical voids, parallel to the time axis and $3$-dimensionally spherical
in cross-section (or that we're in a Lorentz frame in which the voids look
that way).  The fractal is now the distribution of cross-section $3$-spheres
in position space; $\rho$ and $V$ now refer directly to that distribution.
(Voids parallel to the time axis also guarantees time-translation invariance
and therefore Hamiltonian quantum dynamics and unitarity.)

As before, we want to demonstrate that the fractal has the effect of
multiplying the Lorentz-invariant free-space propagator by an expression
similar to \eqref{eq5}.  We can confine the demonstration to the vicinity of the
light cone, since that's the only region where the free-space propagator,
$-q(\abs{x}^2+ie)^{-1}$ with infinitesimal $e$, really matters.  (Also, we assume the
scalar field is massless because we're only interested in short distances.)
Near the light cone, propagation past a $4$-cylinder looks like a plane wave
passing a polarizable $3$-sphere.  And as long as the width of the plane-wave
pulse $\gg$ sphere separation, the basic logic of the dielectric model in
Section~\ref{sec3} still applies, leading again to the multiplier \eqref{eq5},
because scattered fields in the near field exactly reproduce statically
induced dipoles (see for example \cite[Sec.~9.2]{jack}).

If plane-wave pulse-width not $\gg$ sphere separation, then presumably scattered
waves from nearby spheres are unable to add coherently, in which case one
can't include the factor $\Phi_k$ for that separation.  In this way position-space
power-law screening \eqref{eq5} is augmented by an extra momentum-space factor
\begin{equation}
\left(\frac{\omega}{\rho^{1/3}}\right)^{\!+\ln(1-\rho V)/\ln\xi}
\label{eq6}
\end{equation}
where $\omega$ is the source frequency.

\section{Discussion}\label{sec5}

By focusing on scalar fields, we have begun a longer-term attempt to provide
an explicit physical basis for dimensional regularization.  In this paper,
dimensional regularization emerges as a considerable idealization: It ignores
non-unity $f(\Omega)$ or $g(\Omega)$ for small $\varepsilon$; and fractal screening
(expression \eqref{eq4}) is really stepwise, not literally a smooth power law
(although perhaps the steps can be eliminated by defining the fractal in the
limit of vanishing $\ln\xi$ and $\rho V$ with finite ratio).  These non-idealizations
clearly depend on details of how the underlying fractal is defined.

We readily acknowledge weaknesses in our reasoning.  In particular, it hinges
on various approximations and idealizations, including a reliance on spherical
voids or cross sections, dipole-only responses, and multiplicatively iterative
dielectric calculations.

Generalization to non-scalar fields is by no means assured, since they
involve not just power-law screening but also nontrivial component index
structure and constraints related to gauge invariance.

But if the fractal construction really does generalize to all types of fields
(and if it also generalizes to curved geometries), then one can speculate
that literally setting spacetime's dimension to $4-\varepsilon$ might render
gravity's renormalizability a non-issue without unification with other
forces or as-yet unobserved symmetries (assuming nothing discontinuous but
essential happens at $\varepsilon=0$).  Such a scenario has some numerical
plausibility: Consider quantum corrections to the Einstein--Hilbert
Lagrangian $(1/2\kappa^2)(-g)^{1/2}R$, where $\kappa^2=8\pi G/c^4$ is proportional to
Newton's constant; $g$ is the determinant of the metric tensor; and $R$ is the
Ricci scalar, essentially a sum of curvature components.  Assume gravity is
minimally coupled to matter fields but not otherwise unified with matter.
The simplest induced nonrenormalizable interactions, from coupling to a
massless scalar field \cite{hoof74} or massless photons \cite{dese}, are in one loop and
proportional to $(1/\varepsilon)(-g)^{1/2}Q$, where $Q$ is quadratic in curvature
components.  Dimensionally, the generic proportionality constant can only be a
geometric-combinatoric number times $L_P^2/2\kappa^2$, where $L_P$ is the Planck length.
The tightest ``fifth force'' observational bound \cite{berr} on $R+a_2R^2$ extensions
of the Einstein--Hilbert action (i.e.\ $Q=R^2$) is $a_2<4\times 10^{-9}$\,m$^2$, suggesting
$\varepsilon= L_P^2/a_2 > 10^{-61}$ (ignoring geometric and combinatoric factors),
easily small enough to have escaped observation.  This echoes an earlier
suggestion \cite{cran} that gravity's non-renormalizability could be mitigated with
a self-similar distribution of virtual black holes.

\begin{acknowledgements}
I acknowledge helpful correspondence with S. Deser,
S. Libby and C.K. Zachos.
\end{acknowledgements}

\end{document}